\documentclass[conference, a4paper]{APSIPA2021}
\usepackage[psamsfonts]{amssymb}
\usepackage{amsmath,graphicx,booktabs,}
\usepackage{amsmath}
\usepackage{diagbox, makecell}
\usepackage{multirow}
\usepackage{colortbl}
\usepackage{lipsum}
\usepackage{threeparttable}

\begin{document}
	
	\title{DMF-Net: A decoupling-style multi-band fusion model for full-band speech enhancement}

\author{%
	\authorblockN{%
		Guochen Yu\authorrefmark{1}\authorrefmark{2}, Yuansheng Guan\authorrefmark{2}\authorrefmark{3},  Weixin Meng\authorrefmark{2}, Chengshi Zheng\authorrefmark{2}, Hui Wang\authorrefmark{1}, and
		Yutian Wang\authorrefmark{1}
	}
	\authorblockA{%
		\authorrefmark{1}
		State Key Laboratory of Media Convergence and Communication, Communication University of China, Beijing, China \\
		E-mail: \{yuguochen, wangyutian, hwang\}@cuc.edu.cn }
	\authorblockA{%
		\authorrefmark{2}
		Key Laboratory of Noise and Vibration Research, Institute of Acoustics, Chinese Academy of Sciences, Beijing, China\\
		E-mail: \{mengweixin, cszheng\}@mail.ioa.ac.cn }
	\authorblockA{%
	\authorrefmark{3}
	School of Electronics and Communication Engineering, Guangzhou University, Guangzhou, China\\
	E-mail: 2111907059@e.gzhu.edu.cn }
}

\maketitle
\thispagestyle{empty}		

		\begin{abstract}
For the difficulty and large computational complexity of modeling more frequency bands, full-band speech enhancement based on deep neural networks is still challenging. Previous studies usually adopt compressed full-band speech features in Bark and ERB scale with relatively low frequency resolution, leading to degraded performance, especially in the high-frequency region. In this paper, we propose a decoupling-style multi-band fusion model to perform full-band speech denoising and dereverberation. Instead of optimizing the full-band speech by a single network structure, we decompose the full-band target into multi sub-band speech features and then employ a multi-stage chain optimization strategy to estimate clean spectrum stage by stage. Specifically, the low- (0-8 kHz), middle- (8-16 kHz), and high-frequency (16-24 kHz) regions are mapped by three separate sub-networks and are then fused to obtain the full-band clean target STFT spectrum. Comprehensive experiments on two public datasets demonstrate that the proposed method outperforms previous advanced systems and yields promising performance in terms of speech quality and intelligibility in real complex scenarios. 
		\end{abstract}

		\section{Introduction}

		Speech enhancement (SE) aims to extract clean speech target from the noise-corrupted mixture, so as to improve speech quality~{\cite{loizou2013speech}} in real-time speech communication scenarios. Over the past few years, a multitude of SE algorithms based on deep neural networks (DNNs) have demonstrated their superior capability in dealing with non-stationary noise under low signal-to-noise ratio (SNR) conditions~{\cite{wang2018supervised}}. However, due to the high computational complexity caused by a large number of neurons and weights, most of these existing DNN-based SE approaches focus on narrow- or wide-band signals at the sampling rate of 8 or 16 kHz. For the difficulty of modeling the higher-dimensional features, it is still challenging to implement super-wideband (32 kHz) and full-band (48 kHz) SE. 
				
		In this regard, instead of using the Fourier spectrum directly, lower-dimensional compressed features have already been considered as inputs for deep learning-based full-band SE approaches~{\cite{valin2018hybrid, valin2020perceptually, schroter2021deepfilternet}}. In~{\cite{valin2018hybrid}}, Bark-scale spectrum with 22-dimensional Bark-frequency cepstral coefficients (BFCC) was adopted as input features and 22 ideal critical band gains were mapped, which can reduce the model size and computational complexity simultaneously. More recently, based on the human perception of speech signals, PercepNet developed a perceptual band representation with 32 triangular spectral bands~{\cite{valin2020perceptually}}, spaced according to the human hearing equivalent rectangular bandwidth (ERB). Obviously, the frequency resolution of the spectrum in Bark scale and that in ERB scale are much lower than the Fourier spectrum, leading to the leakage of information among frequency bands. In~{\cite{lv2021s}}, DCCRN~{\cite{hu2020dccrn}} was extended for super-wideband SE by simultaneous modeling of sub-band and full-band (SAF) features, in which the input features are the 257-dimensional Fourier spectrum, achieving superior performance when compared with other state-of-the-art (SOTA) full-band SE approaches~{\cite{lv2021s}}.

		Motivated by the curriculum learning, decoupling-style phase-aware methods have thrived in the SE area, where the original complex-spectrum estimation problem is decomposed into two sub-stages, i.e., magnitude and phase~{\cite{li2021two, li2021simultaneous, yu2021dual,yu2022dbt}}. Specifically, only the magnitude estimation is involved in the first stage, followed by the complex spectrum refinement in the second stage. This paper proposes a decoupling-style multi-band fusion model, dubbed DMF-Net for full-band SE step by step. To be specific, we adopt a two-step strategy which consists of a pre-trained complex-domain-based SE network (LF-Net) for the low-frequency band (0-8 kHz) and two magnitude-based SE networks (MF-Net and HF-Net) for the middle- and high-frequency bands (8-16 and 16-24 kHz). In LF-Net, inspired by the preliminary study~{\cite{ li2021simultaneous}}, we employ a decoupling-style strategy for wide-band speech enhancement by three sub-networks. In the first two sub-networks, the optimization \emph{w.r.t.} denoising and dereverberation are serially implemented in the magnitude domain. \emph{i.e.}, we decouple the original complex multi-target optimization into magnitude and phase, and only focus on magnitude estimation. Afterward, the phase information can be effectively recovered based on previous results by introducing a global residual connection. Then, we couple the pretained LF-Net with two higher-frequency refinement networks, namely MF-Net and HF-Net, to denoise the 8-16 kHz and 16-24 kHz regions. Note that the estimated low-frequency features are also fed into MF-Net and HF-Net to provide extra guidance. Finally, we fuse the low-, middle- and high-frequency regions to obtain the full-band speech, which is then fed into a low-complexity post-processing module to further suppress the residual noise~{\cite{ ke2021low}}.
		Experimental results on two public datasets show that DMF-Net achieves remarkable performance and outperforms SOTA super-wideband and full-band baselines.
		
		The remainder of the paper is organized as follows. In Section~{\ref{Sec2}}, the proposed framework is described in detail. The experimental setup is presented in Section~{\ref{Sec3}}, while Section~{\ref{Sec4}} gives the results. Some conclusions are drawn in Section~{\ref{Sec5}}.

		\begin{figure*}[t]
			\centering
			\includegraphics[width=1.95\columnwidth]{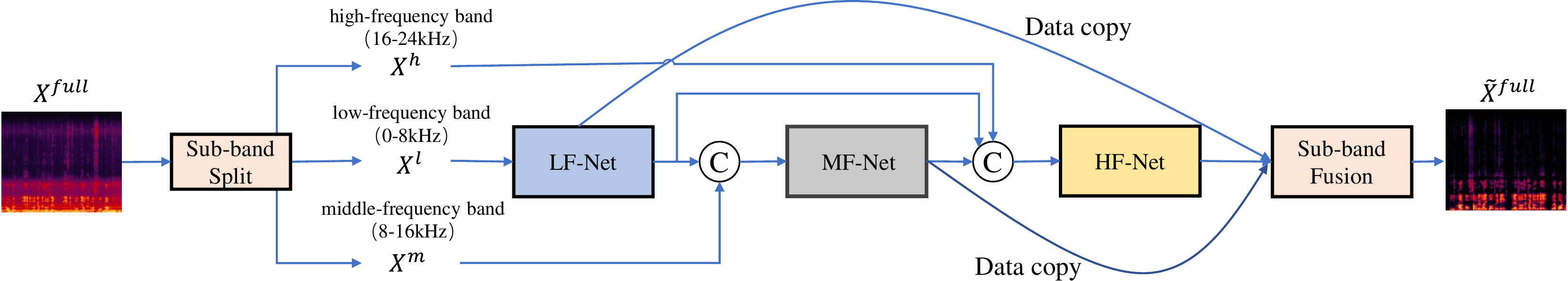}

			\caption{ The overall diagram of the proposed system. $\textcircled{c}$ denotes the concatenation operation in the channel axis.}
			\label{fig:diagram-system}

		\end{figure*}

	\begin{figure}[t]
	\centering
	\includegraphics[width=0.95\columnwidth]{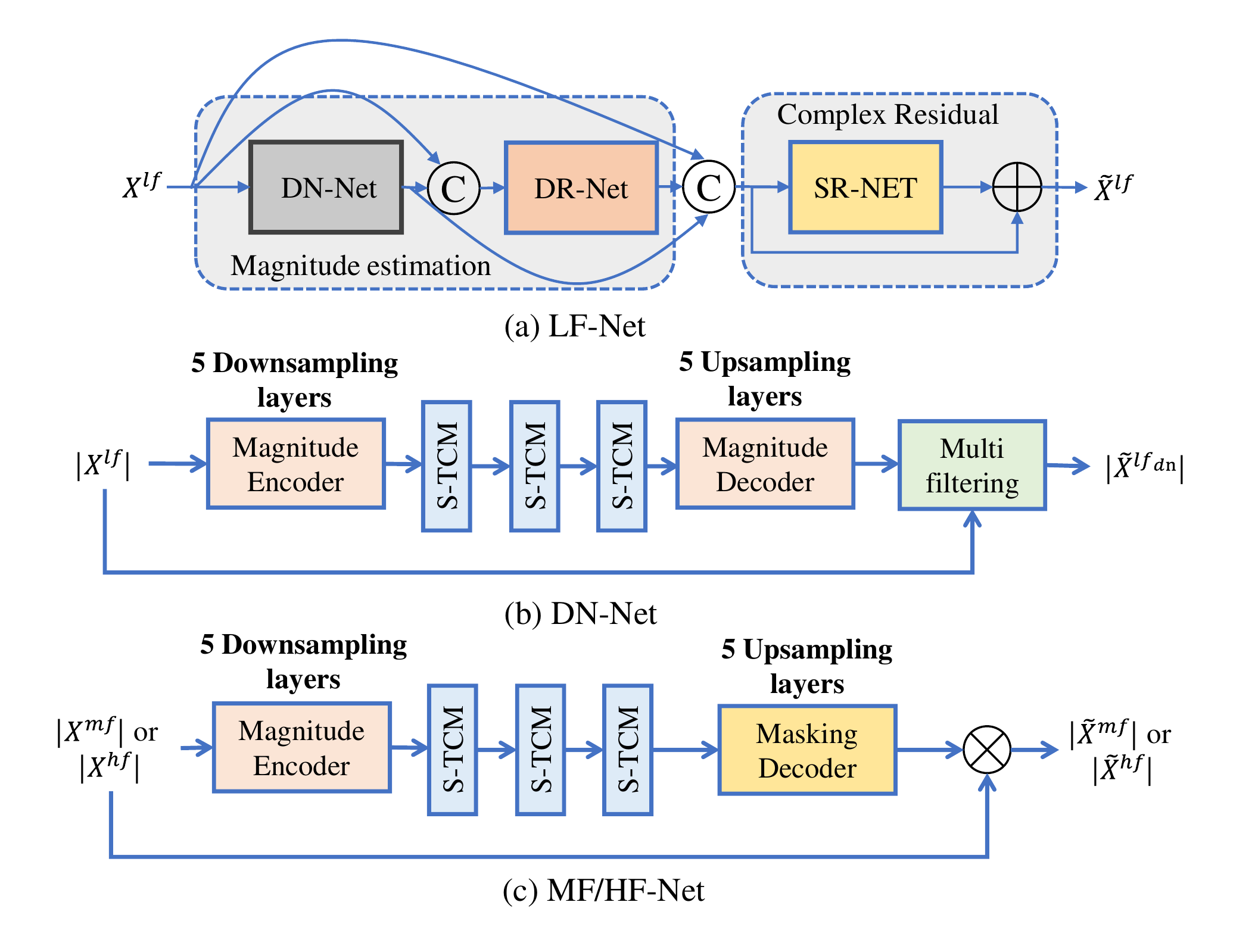}

	\caption{ The detailed network architecture of DMF-Net. (a) The diagram of LF-Net. (b) The diagram of DN-Net in LF-Net. (c) The diagram of MF/HF-Net.}
	\label{fig:networks}

	\end{figure}

		\section{Methodlogy\label{Section2}}
		\label{Sec2}
		\subsection{ Decoupling-style low-frequency band SE \label{Section23}}
		Because human speech contains lots of harmonics and provides more information for the frequency ranging from 0 to 8 kHz, we first employ a low-frequency region estimation sub-network, dubbed LF-Net, to remove the noise and late reverberation in the complex domain. Similar to the previous work~{\cite{li2021simultaneous}}, LF-Net consists of three sub-networks, namely a denoising network (DN-Net), a dereverberation network (DR-Net), and a spectral refinement network (SR-Net), as illustrated in Figure~{\ref{fig:networks}}(a). Inspired by the effectiveness of decoupling-style SE methods~{\cite{li2021two, li2021simultaneous, yu2021dual}}, we decouple the original spectrum estimation into spectral magnitude and phase, and only the magnitude is processed in DN-Net and DR-Net, without recovering the phase. After denoising and dereverberation, we couple the coarsely processed magnitude and the original noisy phase to obtain the coarse-estimated complex spectrum. For SR-Net, it receives both noisy and coarse-estimated complex spectra, and further refines the complex spectrum details as well as restores the clean phase. Note that different from explicitly estimating the whole complex spectrum, we introduce a global residual connection to only focus on estimating the residual missing details of complex spectra, which alleviates the overall burden of the network. The whole procedure of LF-Net can be given by:

		\begin{gather}		
			\label{eqn5}
			|\tilde{X}^{lf_{dn}}| = \mathcal{G}_{dn}\left( |X^{lf}|; \Phi_{1} \right),\\
			|\tilde{X}^{lf_{dr}}| = \mathcal{G}_{dr}\left( |\tilde{X}^{lf_{dn}}|, |X^{lf}|; \Phi_{2} \right),\\
			\tilde{X}^{lf_{dn}} = |\tilde{X}^{lf_{dn}}| \exp(j\theta_{X^{lf}}), \tilde{X}^{lf_{dr}} = |\tilde{X}^{lf_{dr}}| \exp(j\theta_{X^{lf}}),\\
			\tilde{X}^{lf_{sr}} = \tilde{X}^{lf_{dr}} + \mathcal{G}_{sr}\left(\tilde{X}^{lf_{dn}}, \tilde{X}^{lf_{dr}}, X^{lf}; \Phi_{3} \right)
		\end{gather}
		where $\left( \tilde \cdot \right)^{lf_{dn}}$, $\left( \tilde\cdot \right)^{lf_{dr}}$ and $ \left( \tilde\cdot \right)^{lf_{sr}}$ denote the outputs of DN-Net, DR-Net and SR-Net, respectively. $|X^{lf}|$ denotes the noisy low-frequency spectral magnitude. $\mathcal{G}_{dn}$, $\mathcal{G}_{dr}$, and $\mathcal{G}_{sr}$ denote the mapping functions of the corresponding three modules with parameter set $\Phi_{\left(\cdot\right)}$. $\theta_{X^{lf}}$ denotes the noisy phase of low-frequency bands.
		
		For DN-Net and DR-Net, similar to~{\cite{li2021simultaneous}}, we employ the multi-frame (MF) filtering to capture the correlations between neighboring frames~{\cite{mack2019deep}} in the magnitude domain, which can be given by:

		\begin{gather}
			\label{eqn7}
			|\tilde{X}_{l, :}^{lf_{dn}}| = \sum_{\tau=1}^{k} \tilde{M}^{lf_{dn}}_{\tau, :, :} |X^{lf}_{l-\tau, :}|,\\ 
			|\tilde{X}_{l, :}^{lf_{dr}}| = \sum_{\tau=1}^{k} \tilde{M}^{lf_{dr}}_{\tau, :, :} |\tilde{X}^{lf_{dn}}_{l-\tau, :}|,
		\end{gather}		
		where $l$ and $k$ denote the frame index and the filter length, respectively, and we set $k$ to 5 in this paper. 
				
		\subsection{Masking-basked middle/high-frequency bands SE\label{Section21}}
		After optimizing the low-frequency band, we employ two sub-networks, dubbed MF-Net and HF-NET, to refine the middle-frequency (8-16 kHz) and high-frequency (16-24 kHz) regions stage by stage, as illustrated in Figure~{\ref{fig:diagram-system}}. Based on the fact that speech in higher frequency bands tends to contain lower energies and fewer harmonics, we only estimate the spectral gain in the magnitude domain and retain the phase unaltered. This can reduce the complexity of the whole network and evade the implicit compensation effect between magnitude and phase estimation. To introduce explicit information interaction between low- and high-frequency components, the output of LF-Net is also fed into MF-Net and HF-Net. To be specific, we concatenate the estimated spectral magnitude of LF-Net with the original noisy middle-frequency spectral magnitude in the channel axis as the input of MF-Net, while the noisy high-frequency spectral magnitude is concatenated with the output spectral magnitude of LF-Net and MF-Net in the channel axis. In a nutshell, the whole procedure of the middle- and high-frequency bands modeling can be formulated as:
		\begin{gather}
			\label{eqn6}
			|\tilde{X}^{mf}| = \lvert{X^{mf}}\rvert \otimes \mathcal{G}_{mf}\left( |X^{mf}|; \Phi_{4} \right),\\
			|\tilde{X}^{hf}| = \lvert{X^{hf}}\rvert \otimes \mathcal{G}_{hf}\left( |X^{hf}|; \Phi_{5} \right),\\
			\tilde{X}^{mf} = |\tilde{X}^{mf}| \exp(j\theta_{X^{mf}}), \tilde{X}^{hf} = |\tilde{X}^{hf}| \exp(j\theta_{X^{hf}}),
		\end{gather} 	
		where $|\tilde{X}^{mf}|$ and $|\tilde{X}^{hf}|$ denote the outputs of MF-Net and HF-Net, respectively. $\mathcal{G}_{mf}$ and $\mathcal{G}_{hf}$ denote the masking-based functions of MF-Net and HF-Net with parameter set $\Phi_{\left(\cdot\right)}$. $\theta_{X^{mf}}$ and $\theta_{X^{hf}}$ denote the noisy phase of middle- and high-frequency bands, respectively. After estimating three sub-band features, we employ a sub-band fusion operation to obtain the full-band spectrum by simply stacking the low-, middle- and high-frequency bands along the frequency axis. Note that we average the overlapped bands in the middle- and high-frequency bands.
		
		\subsection{Network architecture \label{Section22}}
		
		Inspired by the promising performance of temporal convolutional modules (TCMs) for long-term sequence modeling in the SE area~{\cite{pandey2019tcnn, luo2019conv}}, we insert cascaded TCMs in a typical convolutional encoder-decoder topology~{\cite{zhao2020noisy}}. For LF-Net, the diagram of DN-Net and DR-Net is shown in Figure~{\ref{fig:networks}}(b), in which SR-Net utilizes the same encoder as DN-Net and two decoders to recover both real and imaginary (RI) components. Taking DN-Net as an example, the encoder is comprised of five downsampling blocks, each of which is composed of a convolutional layer, InstanceNorm (IN), and PReLU, with kernel size being (2, 3) in the time and frequency axis except (2, 5) in the first block. The number of channels is set to 64, and the stride is set to (1, 2) to halve the frequency dimension. For the decoder, it employs a similar architecture to the encoder except for using the deconvolutional layers to recover the original size. To mitigate the information loss during the encoding process, skip connections are utilized in each block of the encoder-decoder. One can find that we keep the time resolution unaltered to guarantee causal implementation. To reduce the number of parameters, we utilize a lightweight sequential modeling unit called squeezed TCM (S-TCM) similar to the previous study~{\cite{li2021two, li2021simultaneous}}, where the feature size is first compressed into 64 rather than 512 as in~{\cite{luo2019conv}}, followed by dilated convolutions. For each sub-network in LF-Net, we utilize three groups of TCMs, each of which stacks 6 S-TCMs with dilation rate exponentially increasing to obtain a large temporal receptive field, \emph{i.e.}, $d = \{1, 2, 4, 8, 16, 32\}$. 
		
		For MF-Net and HF-Net, we employ the same encoder and S-TCMs as those in LF-Net, and the only difference is that the dual-path masking-based decoder is adopted in both MF-Net and LF-Net. Similar to~{\cite{yu2021dual,yu2022dbt}}, a dual-path mask module is operated to obtain the magnitude gain function by a 2-D convolution and a dual-path tanh/sigmoid nonlinearity operation, followed by a 2-D convolution and a sigmoid activation function.

		\subsection{Loss function \label{Section24}}
		In DMF-Net, we adopt a two-stage training pipeline for 0-8 and 8-24 kHz sub-bands. Firstly, we train LF-Net with a multi-stage paradigm until convergence. This is to say, before training the network of the current stage, we pre-train the network of the last stage and then freeze the weights. Specifically, DN-Net and DR-Net adopt MSE loss in the magnitude domain while SR-Net calculates loss functions on the RI components and the magnitude of the estimated spectrum, which can be expressed as:				
		\begin{gather}	
			\mathcal{L}_{lf_{dn}}=\left \| \lvert \tilde X^{lf_{dn}} \rvert - \lvert S^{lf_{dn}} \rvert \right \|^{2}_{F},\\		
			\mathcal{L}_{lf_{dr}}=\left \| \lvert \tilde X^{lf_{dr}} \rvert - \lvert S^{lf} \rvert \right \|^{2}_{F},\\			
			\mathcal{L}_{lf_{sr}}^{Mag}=\left \| \lvert \tilde X^{lf_{sr}} \rvert - \lvert S^{lf} \rvert \right \|^{2}_{F},\\
			\mathcal{L}_{lf_{sr}}^{RI}=\left \|\widetilde{X}^{lf_{sr}}_r-S^{lf}_r \right \|_{F}^2 +\left \|\widetilde{X}^{lf_{sr}}_i-S^{lf}_i \right \|_{F}^2,\\
			\mathcal{L}_{lf_{sr}}=\mu \mathcal{L}_{lf_{sr}}^{RI}+(1-\mu ) \mathcal{L}_{lf_{sr}}^{Mag} ,
		\end{gather}
		where $\mathcal{L}_{lf_{dn}}$, $\mathcal{L}_{lf_{dr}}$ and $\mathcal{L}_{lf_{dn}}$ denote the loss functions for DN-Net, DR-Net and SR-Net, respectively. Note that $\lvert S^{lf_{dn}} \rvert$ denotes the target denoised spectral magnitude for DN-Net with reverberation unaltered and $\lvert S^{lf}\rvert$ denotes the final target spectral magnitude. $S^{lf}_r$ and $S^{lf}_i$ represent the target RI components. With the internal trial, we empirically set $\mu= 0.5$ in all the following experiments.
		
		In the second stage, we freeze the pretained LF-Net and optimize the spectral magnitude of the middle- and high-frequency bands ranging from 8 to 24 kHz by training MF-Net and HF-Net jointly, and the overall loss can be given by:
		\begin{gather}
		\mathcal{L}_{mf}=\left \| \lvert \tilde X^{mf} \rvert - \lvert S^{mf} \rvert \right \|^{2}_{F},\\		
		\mathcal{L}_{hf}=\left \| \lvert \tilde X^{hf} \rvert - \lvert S^{hf} \rvert \right \|^{2}_{F},\\
		\mathcal{L}_{full}=\alpha \mathcal{L}_{mf} + (1-\alpha) \mathcal{L}_{hf}
		\end{gather}
		where $\mathcal{L}_{mf}$ and $\mathcal{L}_{hf}$ denote the loss function for MF-Net and HF-Net, while $\mathcal{L}_{full}$ represent the full loss function of the second stage. $\lvert \tilde X^{mf} \rvert$ and $\lvert \tilde X^{hf} \rvert$ denote the outputs of MF-Net and HF-Net, respectively, while $\lvert \tilde X^{mf} \rvert$ and $\lvert \tilde X^{hf} \rvert$ denote the clean spectral magnitude of middle- and high-frequency bands. We empirically find that $\alpha= 0.5$ suffices in our evaluation.
		
\begin{figure*}[t]
	\centering
	\centerline{\includegraphics[width=2\columnwidth]{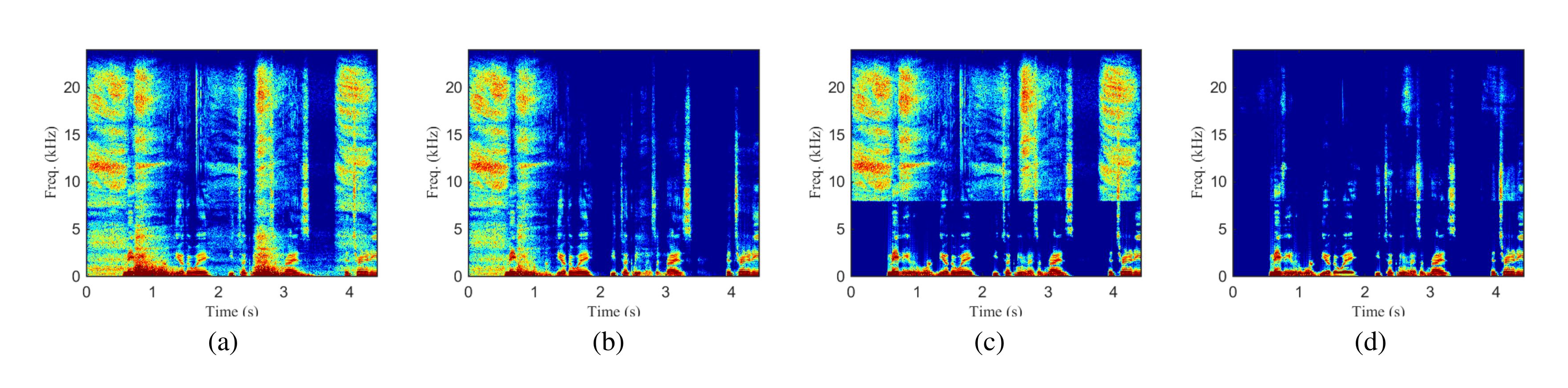}}
	\vspace{-0.3cm} 
	\caption{Visualization of spectrograms of the noisy and different enhanced speech signals. (a) Noisy speech. (b) Enhanced speech by NSNet2. (c) Enhanced speech by LF-Net. (d) Enhanced speech by DMF-Net.  }
	\label{fig:visualization}
	\vspace{-0.3cm} 
\end{figure*}

		\section{Experiments\label{Section3}}
		\label{Sec3}

		\subsection{Datasets\label{Section31}}

		To evaluate the performance of our framework, we conduct extensive experiments on two public datasets, namely VoiceBank + DEMAND dataset~\cite{valentini2016investigating} and the 4th DNS-Challenge dataset~\cite{dubeyicassp}, where all the utterances are sampled at 48 kHz. 
		
		\textbf{VoiceBank + DEMAND}: The dataset is a selection of the VoiceBank corpus~{\cite{veaux2013voice}} with 28 speakers for training and another 2 unseen speakers for testing. The training set includes 11,572 noisy-clean pairs, while the test set contains 824 pairs. For the training set, the audio samples are mixed with one of the 10 noise types, including two artificial noise processes, \emph{i.e.}, babble and speech shaped noise, and eight real recording noise processes taken from the Demand database~{\cite{thiemann2013diverse}}. The test utterances are created with 5 unseen noises taken from the Demand database at SNRs of $\left\{2.5\rm{dB}, 7.5\rm{dB}, 12.5\rm{dB}, 17.5\rm{dB}\right\}$.
		
		\textbf{DNS-Chanllenge}: It consists of a wide range of clean speech, noise clips, and RIRs, which simulate complicated acoustic scenarios. Clean speech consists of six languages including English, French, German, Italian, Russian and Spanish. For this dataset, we totally generate around 600 hours of noisy-clean pairs. To be specific, we synthesize the noisy sets with around 350 hours for the English language. For the other languages, around 150 hours of training sets are generated (30h for each language). Besides, we generate the rest of 100 hours of training sets with VCTK corpus~{\cite{veaux2013voice}} and emotional utterances. To generate reverberant noisy training data, we use 248 real and 60,000 synthetic room impulse responses (RIR) from openSLR26 and openSLR28 datasets~{\cite{ko2017study}}. The direct and early reflected speech components are chosen as the training target when considering that only late reverberation degrades speech quality/intelligibility~{\cite{zhao2020monaural}}. During each mixing process, the clean speech is convolved with a randomly selected RIR, and is then mixed with the noise with the SNR range of $\left(-5\rm{dB}, 15\rm{dB}\right)$. For the testset, we utilize the 4th DNS-Challenge blind set{\footnote{https://github.com/microsoft/DNS-Challenge}}, which consists of 859 real clips, each with a duration of 10s.

		\subsection{Implementation setup\label{Section32}}
		
		The 20ms Hanning window is utilized, with 50\% overlap between adjacent frames. To extract the features, 960-point FFT is utilized and 481-dimension spectral features are obtained. Due to the efficacy of power compression in both dereverberation and denoising tasks~{\cite{li2021importance}}, we conduct the compression toward the spectral magnitude before sending into the network while remaining the phase unaltered, and the compression parameter is $\beta=0.5$, which is optimal or near-optimal. 
		All the models are optimized using Adam($\beta_{1}=0.9$, $\beta_{2}=0.999$)~{\cite{kingma2014adam}}. In the training process, we employ a two-stage training paradigm for better performance. In the first stage, we pretrain LF-Net with the initialized learning rate (LR) is set to 1e-3 on the INTERSPEECH 2021 DNS Challenge datasets~{\cite{reddy2021interspeech}}, where all the utterances are sampled at 16000 hz. In the second stage, we jointly train LF-Net with MF-Net and HF-Net as the proposed DMF-Net with the learning rate of 5e-4. The batch size is set to 16 at the utterance level.
		
		\renewcommand\arraystretch{1.2}
	\begin{table}[!h]
	\centering
	\footnotesize
	\setlength\tabcolsep{2pt}
	\caption{Results of various models on Voice Bank and DEMAND set. "$-$" denotes that the result is not provided in the original paper.}
	\resizebox{0.95\columnwidth}{!}{
		\begin{tabular}{l|c|c|ccccc}
			\toprule
			Models   &year & Para.(M) & PESQ & STOI(\%) &CSIG & CBAK & COVL \\
			\midrule
			Noisy       &  -  & -  & 1.97 & 92.1 & 3.35 & 2.44 & 2.63   \\
			RNNoise~{\cite{valin2018hybrid}}      & 2020  & 0.06  & 2.34 & 92.2 & 3.40 & 2.51 & 2.84   \\
			GCRN (full)~{\cite{tan2019learning}}        & 2019   & 10.59  & 2.71 & 93.8 & 4.12 & 3.23  & 3.41      \\
			PercepNet~{\cite{valin2020perceptually}}    & 2020     & 8.00  & 2.73  & - & -    & - &-  \\
			DCCRN~{\cite{hu2020dccrn}}        & 2020   & 3.70  & 2.54 & 93.8 & 3.74 & 2.75  & 3.13      \\
			DeepFilterNet~{\cite{schroter2021deepfilternet}} & 2021  & 1.80  & 2.81  & - & -    & - &- \\
			{S}-DCCRN~{\cite{lv2021s}}      & 2022  & 2.34  & 2.84  & 94.0 & 4.03 & 2.97 & 3.43 \\
			CTS-Net (full)~{\cite{li2021two}}        & 2020   & 7.09  & 2.92  & 94.3 & 4.22 & 3.43 & 3.62     \\
			DMF-Net (\textbf{Pro.})      & 2022  & 7.84  & \textbf{2.97} & \textbf{94.4}  & \textbf{4.26} & \textbf{3.52} & \textbf{3.62} \\
			~ w/o SR-Net (\textbf{Pro.})      & 2022  & 5.45  & 2.85 & 94.1  & 4.08 & 3.23 & 3.34  
			\\\bottomrule
	\end{tabular}}		
	\label{tab:vctk}
	
\end{table}

\renewcommand\arraystretch{0.95}
\begin{table}[th]
		\vspace{-0.5cm} 
	\setlength{\abovecaptionskip}{2pt}
	\caption{Subjective evaluation on DNS challenge blind test set.}
	\label{tab:dns_sub}
	\setlength{\tabcolsep}{2.81mm}
	\centering
	\begin{tabular}{cp{30pt}p{50pt}p{30pt}l}
		\toprule
		\makecell[c]{Model} & \makecell[c]{SIG\\MOS} & \makecell[c]{BAK\\MOS} & \makecell[c]{OVR\\MOS} &  \\
		\midrule
		Noisy          & \makecell[c]{4.29}                & \makecell[c]{2.15}                           & \makecell[c]{2.63}                 &  \\
		NSNet2         & \makecell[c]{3.62}                & \makecell[c]{3.93}                          & \makecell[c]{3.26}                 &  \\
		DMF-Net       & \makecell[c]{\textbf{3.86}}       & \makecell[c]{\textbf{4.47}}                 & \makecell[c]{\textbf{3.66}}        &  \\
		\bottomrule
	\end{tabular}

\end{table}

\renewcommand\arraystretch{1}
\begin{table}[t!]
	\centering
	\footnotesize
	\caption{DNSMOS P.835 and P.808 results on DNS-2022 blind test set.}
	\scalebox{1}{
		\begin{threeparttable}
			
			\begin{tabular}{lcccc}
				\toprule
				\multirow{2}*{Model}   &\multicolumn{3}{c}{DNSMOS P.835$^*$} &{DNSMOS$^*$}\\
				\cline{2-4}
				& SIG& BAK &OVRL &P.808\\
				
				\midrule
				Noisy     & 4.14 & 2.94 & 3.27 &3.03   \\
				NSNet2~{\cite{braun2020data}}  & 3.87 & 4.21 & 3.58 &3.57 \\
				DMF-Net  & 3.92 & \textbf{4.57} & \textbf{3.72} & \textbf{3.61} \\

				\bottomrule
			\end{tabular}
			\begin{tablenotes}
				\item *: Calculated on downsampled speech at 16000 Hz.  
			\end{tablenotes}
	\end{threeparttable}}
	\label{tab:dnsmos}
\end{table}

		\vspace{-0.3cm} 
		\section{Experimental results and discussion\label{Section4}}
		\label{Sec4}		
		
		We use the perceptual evaluation of speech quality (PESQ)~{\cite{rix2001perceptual}}, short-time objective intelligibility (STOI)~{\cite{taal2010short}}, CSIG, CBAK, and COVL~{\cite{hu2007evaluation}} to evaluate speech enhancement performance. Higher values indicate better performance. 
		
		\subsection{Comparison with SOTA methods on Voice-Bank dataset\label{Section41}}
		
		First, we conduct extensive experiments on VoiceBank + DEMAND to compare the proposed methods with several state-of-the-art (SOTA) full-band and super-wideband SE methods, including RNNoise~{\cite{valin2018hybrid}}, GCRN (full-band version)~{\cite{tan2019learning}}, PercepNet~{\cite{valin2020perceptually}}, DCCRN~{\cite{hu2020dccrn}} (super-wideband version), DeepFilterNet~{\cite{schroter2021deepfilternet}}, S-DCCRN~{\cite{lv2021s}} and CTS-Net (full-band version)~{\cite{li2021two}}. Note that we re-implement GCRN (full) and CTS-Net (full) with two more downsampling-upsampling layers to conduct full-band speech enhancement, which are also trained with power compression for a fair comparison. For other baselines, we directly use the reported results from their original papers. We also conduct ablation study to show the importance of phase recovery for the low-frequency band (i.e., w/o SR-Net), which removes the complex spectral refinement in LF-Net. Due to the lack of the full-band PESQ evaluation tool, we resample the outputs of all methods to 16 kHz for a fair comparison when measuring the PESQ.
		From Table~\ref{tab:vctk}, we can have the following observations. Compared with other SOTA full-band methods, the proposed method achieves relatively better performance in terms of all metrics. On average, DMF-Net yields 0.24 PESQ improvement compared to Percepnet with less parameter burden. This demonstrates that the higher resolution spectral features can leverage the information of frequency bins and improve speech performance. Going from the S-DCCRN to DMF-Net, we also observe consistent improvements in all five objective metrics, e.g., average 0.13, 0.4\%, 0.23, 0.55 and 0.19 increase in PESQ, STOI, CSIG, CBAK and COVL scores are achieved. This indicates that the multi-stage strategy can obtain better performance over single-stage
		 complex-domain networks. 
		\subsection{Subjective and DNSMOS comparison on the DNS-Challenge dataset\label{Section42}}
		Then, we evaluate our model with the 4th DNS-Challenge baseline, dubbed NSNet2~{\cite{braun2020data}}, on the 2022 DNS Challenge dataset. In Table~{\ref{tab:dns_sub}}, we present the results with ITU-T P.835~{\cite{naderi2020crowdsourcing}} subjective evaluation scores, which is provided by the DNS-Challenge organizers, including speech quality (SIG), background noise quality (BAK), and overall audio quality (OVRL). With less speech distortion, the proposed approach also yields much better performance in background noise and overall MOS than the baseline. 
				
		Additionally, we also utilize the non-intrusive perceptual speech quality metric, namely DNSMOS P.808~{\cite{reddy2021dnsmos}} and P.835~{\cite{reddy2022dnsmos}}, to evaluate the perceptual speech performance, which are based on ITU-T
		P.808~{\cite{naderi20_interspeech}} and P.835~{\cite{naderi21_interspeech}}, respectively. As shown in Table~{\ref{tab:dnsmos}, compared with NSNet2, a standard baseline system for DNS-2022~{\cite{braun2020data}}, the proposed approach yields consistently better performance in speech distortion (SIG), background noise (BAK) and overall quality (OVRL). For example, compared with NSNet2, DMF-Net provides average 0.05 SIG, 0.36 BAK and 0.14 OVRL improvements in DNSMOS P.835 scores and 0.04 improvement in DNSMOS P.808 score. This verifies the superiority of our decoupling-style approach in suppressing the background noise and recovering speech components in practical acoustic scenarios
			
		Spectrograms of a noisy utterance and its corresponding enhanced utterances by NSNet2, LF-Net and DMF-Net are presented in Figure~{\ref{fig:visualization}}(a)-(d). Focusing on the low-frequency band ranging from 0 to 8 kHz in Figure~{\ref{fig:visualization}}(c), one can see that LF-Net better suppresses background noise than NSNet2, which indicates the effectiveness of the proposed decoupling-style approach in handling the low-frequency band. As illustrated in Figure~{\ref{fig:visualization}}(d), by fusing LF-Net, MF-Net and HF-Net, DMF-Net demonstrates the remarkable performance in reducing the background noise and preserving the speech components for the full-band speech.

		%
		%

		\section{Conclusion\label{Section5}}
		\label{Sec5}

		In this paper, we propose a novel decoupling-style multi-band fusion speech enhancement model running on 48 kHz speech signals. Instead of optimizing the full-band speech by a single network structure, we decompose the full-band target into multi sub-bands, i.e., low- (0-8 kHz), middle- (8-16 kHz) and high- (16-24 kHz) frequency bands, and employ a multi-stage chain optimization strategy to estimate the clean speech target. Specifically, conducting on the STFT domain, we design three sub-networks to respectively focus on each sub-band and adopt a fusion operation to obtain an estimate of the full-band clean spectrum. Experimental results demonstrate that the proposed method achieves state-of-the-art performance in terms of speech quality and intelligibility in both low- and high-frequency regions. 
		
		\bibliographystyle{IEEEbib}
		\bibliography{myrefs}

\begin{thebibliography}{10}

\bibitem{loizou2013speech}
P.~C. Loizou,
\newblock {\em Speech enhancement: theory and practice},
\newblock CRC press, 2013.

\bibitem{wang2018supervised}
D.~L. Wang and J.~Chen,
\newblock ``Supervised speech separation based on deep learning: An overview,''
\newblock {\em IEEE/ACM Trans. Audio. Speech, Lang. Process.}, vol. 26, no. 10,
  pp. 1702--1726, 2018.

\bibitem{valin2018hybrid}
J.~M. Valin,
\newblock ``A hybrid dsp/deep learning approach to real-time full-band speech
  enhancement,''
\newblock in {\em 2018 IEEE 20th international workshop on multimedia signal
  processing (MMSP)}. IEEE, 2018, pp. 1--5.

\bibitem{valin2020perceptually}
J.~M. Valin, U.~Isik, N.~Phansalkar, R.~Giri, K.~Helwani, and A.~Krishnaswamy,
\newblock ``A perceptually-motivated approach for low-complexity, real-time
  enhancement of fullband speech,''
\newblock {\em arXiv preprint arXiv:2008.04259}, 2020.

\bibitem{schroter2021deepfilternet}
H.~Schr{\"o}ter, T.~Rosenkranz, A.~Maier, et~al.,
\newblock ``Deepfilternet: A low complexity speech enhancement framework for
  full-band audio based on deep filtering,''
\newblock {\em arXiv preprint arXiv:2110.05588}, 2021.

\bibitem{lv2021s}
S.~Lv, Y.~Fu, M.~Xing, J.~Sun, L.~Xie, J.~Huang, Y.~Wang, and T.~Yu,
\newblock ``S-dccrn: Super wide band dccrn with learnable complex feature for
  speech enhancement,''
\newblock {\em arXiv preprint arXiv:2111.08387}, 2021.

\bibitem{hu2020dccrn}
Y.~Hu, Y.~Liu, S.~Lv, M.~Xing, and L.~Xie,
\newblock ``Dccrn: Deep complex convolution recurrent network for phase-aware
  speech enhancement,''
\newblock {\em arXiv preprint arXiv:2008.00264}, 2020.

\bibitem{li2021two}
A.~Li, W.~Liu, C.~Zheng, C.~Fan, and X.~Li,
\newblock ``{T}wo {H}eads are {B}etter {T}han {O}ne: {A} {T}wo-{S}tage
  {C}omplex {S}pectral {M}apping {A}pproach for {M}onaural {S}peech
  {E}nhancement,''
\newblock {\em IEEE/ACM Trans. Audio. Speech, Lang. Process.}, vol. 29, pp.
  1829--1843, 2021.

\bibitem{li2021simultaneous}
A.~Li, W.~Liu, X.~Luo, G.~Yu, C.~Zheng, and X.~Li,
\newblock ``A simultaneous denoising and dereverberation framework with target
  decoupling,''
\newblock in {\em Proc. Interspeech}, 2021, pp. 2801--2805.

\bibitem{yu2021dual}
G.~Yu, A.~Li, Y.~Wang, Y.~Guo, H.~Wang, and C.~Zheng,
\newblock ``Dual-branch attention-in-attention transformer for single-channel
  speech enhancement,''
\newblock in {\em Proc. ICASSP}. IEEE, 2022, pp. 7847--7851.

\bibitem{yu2022dbt}
G.~Yu, A.~Li, H.~Wang, Y.~Wang, Y.~Ke, and C.~Zheng,
\newblock ``{DBT-Net}: Dual-branch federative magnitude and phase estimation
  with attention-in-attention transformer for monaural speech enhancement,''
\newblock {\em arXiv preprint arXiv:2202.07931}, 2022.

\bibitem{ke2021low}
Y.~Ke, A.~Li, C.~Zheng, R.~Peng, and X.~Li,
\newblock ``Low-complexity artificial noise suppression methods for deep
  learning-based speech enhancement algorithms,''
\newblock {\em EURASIP Journal on Audio, Speech, and Music Processing}, vol.
  2021, no. 1, pp. 1--15, 2021.

\bibitem{mack2019deep}
Wolfgang Mack and Emanu{\"e}l~AP Habets,
\newblock ``Deep filtering: Signal extraction and reconstruction using complex
  time-frequency filters,''
\newblock {\em IEEE Signal Processing Letters}, vol. 27, pp. 61--65, 2019.

\bibitem{pandey2019tcnn}
A.~Pandey and D.~Wang,
\newblock ``{TCNN}: {T}emporal convolutional neural network for real-time
  speech enhancement in the time domain,''
\newblock in {\em Proc. ICASSP}. IEEE, 2019, pp. 6875--6879.

\bibitem{luo2019conv}
Y.~Luo and N.~Mesgarani,
\newblock ``Conv-tasnet: Surpassing ideal time--frequency magnitude masking for
  speech separation,''
\newblock {\em IEEE/ACM Trans. Audio. Speech, Lang. Process.}, vol. 27, no. 8,
  pp. 1256--1266, 2019.

\bibitem{zhao2020noisy}
Y.~Zhao and D.~Wang,
\newblock ``Noisy-reverberant speech enhancement using denseunet with
  time-frequency attention.,''
\newblock in {\em INTERSPEECH}, 2020, pp. 3261--3265.

\bibitem{valentini2016investigating}
C.~Valentini-Botinhao, X.~Wang, S.~Takaki, and J.~Yamagishi,
\newblock ``Investigating {RNN}-based speech enhancement methods for
  noise-robust text-to-speech,''
\newblock in {\em Proc. SSW}, 2016, pp. 146--152.

\bibitem{dubeyicassp}
H.~Dubey, V.~Gopal, R.~Cutler, A.~Aazami, S.~Matusevych, S.~Braun, S.~E.
  Eskimez, M.~Thakker, T.~Yoshioka, H.~Gamper, et~al.,
\newblock ``Icassp 2022 deep noise suppression challenge,''
\newblock in {\em Proc. ICASSP}. IEEE, 2022.

\bibitem{veaux2013voice}
C.~Veaux, J.~Yamagishi, and S.~King,
\newblock ``The voice bank corpus: Design, collection and data analysis of a
  large regional accent speech database,''
\newblock in {\em Proc. O-COCOSDA/CASLRE}. IEEE, 2013, pp. 1--4.

\bibitem{thiemann2013diverse}
J.~Thiemann, N.~Ito, and E.~Vincent,
\newblock ``The diverse environments multi-channel acoustic noise database: A
  database of multichannel environmental noise recordings,''
\newblock {\em JASA}, vol. 133, no. 5, pp. 3591--3591, 2013.

\bibitem{ko2017study}
T.~Ko, V.~Peddinti, M.~L. Povey, D.and~Seltzer, and S.~Khudanpur,
\newblock ``A study on data augmentation of reverberant speech for robust
  speech recognition,''
\newblock in {\em Proc. ICASSP}. IEEE, 2017, pp. 5220--5224.

\bibitem{zhao2020monaural}
Y.~Zhao, D.~Wang, B.~Xu, and T.~Zhang,
\newblock ``Monaural speech dereverberation using temporal convolutional
  networks with self attention,''
\newblock {\em IEEE/ACM Trans. Audio. Speech, Lang. Process.}, vol. 28, pp.
  1598--1607, 2020.

\bibitem{li2021importance}
A.~Li, C.~Zheng, R.~Peng, and X.~Li,
\newblock ``On the importance of power compression and phase estimation in
  monaural speech dereverberation,''
\newblock {\em JASA Express Letters}, vol. 1, no. 1, pp. 014802, 2021.

\bibitem{kingma2014adam}
D.~Kingma and J.~Ba,
\newblock ``Adam: A method for stochastic optimization,''
\newblock {\em arXiv preprint arXiv:1412.6980}, 2014.

\bibitem{reddy2021interspeech}
C.~KA. Reddy, H.~Dubey, K.~Koishida, A.~Nair, V.~Gopal, R.~Cutler, S.~Braun,
  H.~Gamper, R.~Aichner, and S.~Srinivasan,
\newblock ``Interspeech 2021 deep noise suppression challenge,''
\newblock {\em arXiv preprint arXiv:2101.01902}, 2021.

\bibitem{tan2019learning}
K.~Tan and D.~L. Wang,
\newblock ``Learning complex spectral mapping with gated convolutional
  recurrent networks for monaural speech enhancement,''
\newblock {\em IEEE/ACM Trans. Audio. Speech, Lang. Process.}, vol. 28, pp.
  380--390, 2019.

\bibitem{braun2020data}
S.~Braun and I.~Tashev,
\newblock ``Data augmentation and loss normalization for deep noise
  suppression,''
\newblock in {\em International Conference on Speech and Computer}. Springer,
  2020, pp. 79--86.

\bibitem{rix2001perceptual}
A.~Rix, J.~Beerends, M.~Hollier, and A.~Hekstra,
\newblock ``Perceptual evaluation of speech quality ({PESQ})-a new method for
  speech quality assessment of telephone networks and codecs,''
\newblock in {\em Proc. ICASSP}. IEEE, 2001, vol.~2, pp. 749--752.

\bibitem{taal2010short}
C.~H. Taal, R.~C. Hendriks, R.~Heusdens, and J.~Jensen,
\newblock ``A short-time objective intelligibility measure for time-frequency
  weighted noisy speech,''
\newblock in {\em Proc. ICASSP}. IEEE, 2010, pp. 4214--4217.

\bibitem{hu2007evaluation}
Y.~Hu and P.~C. Loizou,
\newblock ``Evaluation of objective quality measures for speech enhancement,''
\newblock {\em IEEE/ACM Trans. Audio. Speech, Lang. Process.}, vol. 16, no. 1,
  pp. 229--238, 2007.

\bibitem{naderi2020crowdsourcing}
B.~Naderi and R.~Cutler,
\newblock ``A crowdsourcing extension of the {ITU-T} recommendation p. 835 with
  validation,''
\newblock {\em arXiv e-prints}, pp. arXiv--2010, 2020.

\bibitem{reddy2021dnsmos}
C.~K. Reddy, V.~Gopal, and R.~Cutler,
\newblock ``{DNSMOS}: A non-intrusive perceptual objective speech quality
  metric to evaluate noise suppressors,''
\newblock in {\em Proc. ICASSP}. IEEE, 2021, pp. 6493--6497.

\bibitem{reddy2022dnsmos}
C.~K. Reddy, V.~Gopal, and R.~Cutler,
\newblock ``{DNSMOS P. 835}: A non-intrusive perceptual objective speech
  quality metric to evaluate noise suppressors,''
\newblock in {\em Proc. ICASSP}. IEEE, 2022.

\bibitem{naderi20_interspeech}
Babak N. and Ross C.,
\newblock ``{An Open Source Implementation of ITU-T Recommendation P.808 with
  Validation},''
\newblock in {\em Proc. Interspeech 2020}, 2020, pp. 2862--2866.

\bibitem{naderi21_interspeech}
Babak N. and Ross C.,
\newblock ``{Subjective Evaluation of Noise Suppression Algorithms in
  Crowdsourcing},''
\newblock in {\em Proc. Interspeech 2021}, 2021, pp. 2132--2136.

\end{thebibliography}

	\end{document}